\documentclass[eqsecnum,aps]{revtex4}
\def\beq{\begin{equation}}
\def\eeq{\end{equation}}
\usepackage{amsmath}
\usepackage{amsfonts}
\usepackage{amssymb}
\usepackage{amscd}
\usepackage[dvips]{graphicx,color}

\def\btheta{\mbox{\boldmath$\theta$}}

\def\brho{\mbox{\boldmath$\rho$}}

\def\x{{\sf x}}

\def\a{{\sf a}}

\def\E{{\sf E}}

\def\X{{\sf X}}

\begin{document}
\title{Poincar\'e Invariant Quantum Mechanics based on Euclidean Green functions}

\author{W. N. Polyzou}
\affiliation{
Department of Physics and Astronomy, The University of Iowa, Iowa City, IA
52242}

\author{Phil Kopp}
\affiliation{
Department of Physics and Astronomy, The University of Iowa, Iowa City, IA 
52242}

\vspace{10mm}

\date{\today}

\begin{abstract}

We investigate a formulation of Poincar\'e invariant quantum
  mechanics where the dynamical input is Euclidean invariant Green
  functions or their generating functional.  We argue that within this
  framework it is possible to calculate scattering observables,
  binding energies, and perform finite Poincar\'e transformations
  without using any analytic continuation.  We demonstrate, using a
  toy model, how matrix elements of $e^{-\beta H}$ in normalizable
  states can be used to compute transition matrix elements for
  energies up to 2 GeV.  We discuss some open problems.
\end{abstract}

\vspace{10mm}


\maketitle






We investigate the possibility of formulating Poincar\'e invariant
quantum models of few-body systems where the dynamical input is given
by a set of Euclidean-invariant Green functions.  This is an
alternative to the direct construction of Poincar\'e Lie algebras on
few-body Hilbert spaces.  In the proposed framework all calculations
are performed using time Euclidean variables, with no analytic
continuation.
  
One potential advantage of the Euclidean approach is that it has a
more direct relation to Lagrangian based field theory models.  One of
the challenges is the construction of a robust class of suitable model
Green functions.  In this paper do not address this problem; 
we assume that this has already been solved and discuss how one can
calculate observables without analytic continuation.

Most of what we propose in not new, it is motivated the reconstruction
theorem of a quantum theory in Euclidean field theory.  The
fundamental work was done by Osterwalder and Schrader
\cite{Osterwalder:1973dx}\cite{Osterwalder:1974tc}.  The approach
illustrated in this work is strongly motivated by Fr\"ohlich's
\cite{Frohlich:1974} elegant solution of the reconstruction problem
using generating functionals.  One of the interesting observations of
Osterwalder and Schrader is that locality is not needed to construct
the quantum theory.

To keep our discussion as simple as possible we assume that we are
given a Euclidean invariant generating functional for a scalar field
theory.  This input replaces the model Hamiltonain.  We assume that
this generating functional is Euclidean invariant, positive,
reflection positive, and satisfies space-like cluster properties.
These requirements are defined below.  The conditions on the
generating functional imply conditions on Green functions in models
based on a subsets of Green functions.

For a scalar field the Euclidean generating functional $Z[f]$ is the 
functional Fourier transform of the Euclidean path measure:
\beq
Z[f] := {\int D_e[\phi] e^{-A[\phi] + i \phi (f) } \over 
\int D_e[\phi] e^{-A[\phi] }} = 
\sum_{n} {(i)^n \over n!} S_n \underbrace{(f,\cdots,f)}_{\mbox{n times}}. 
\eeq 
where $f(\x)= f(\tau,\mathbf{x})$ is a test function in 
four Euclidean space-time variables
and $S_n(\x_1, \cdots ,\x_n)$ is the $n$-point Euclidean Green function.

The generating functional is Euclidean invariant if $Z[f]=Z[f']$
where $f'(\x) = f(E^{-1}(\x-\a))$ where $\x \to E\x +\a$ is a four-dimensional 
Euclidean transformation of the arguments of $f$.

The generating functional is positive if for every finite sequence of 
real test functions $\{f_i\}$ the matrices $E_{ij} = Z[f_i-f_j]$ are 
non-negative. 

The generating functional is reflection positive if for every sequence
$\{f_i\}$, of real test functions with {\it support for positive Euclidean 
time},  the matrices $M_{ij} = Z[f_i - \Theta f_j]$ are non-negative, where 
$(\Theta f)(\tau,\mathbf{x})=f(-\tau,\mathbf{x})$ is Euclidean time
reflection.

The generating functional satisfies space-like cluster properties if
\beq
\lim_{\vert \mathbf{a}\vert  \to \infty} 
\left ( Z[f+g_{\mathbf{a}}] - Z[f] Z[g] \right ) \to 0
\eeq
where
\beq
g_{\mathbf{a}} (\tau,\mathbf{x} ) := g (\tau,\mathbf{x}-\mathbf{a} ).  
\eeq    
These are the primary requirements that are expected of an acceptable
generating functional.

\section{Hilbert Space}

We begin by representing vectors by wave functionals of the form
\beq
B[\phi]= \sum_{j=1}^{N_b} b_j e^{i \phi (f_j)} \qquad
C[\phi]= \sum_{k=1}^{N_c} c_k e^{i \phi (g_k)}
\label{b.1}
\eeq
where $b_j$ and $c_k$ are complex constants and $f_j(\x)$ and $g_k(\x)$ 
are real Euclidean test functions.  The argument ``$\phi$'' plays the role of 
a formal integration variable.

We define a Euclidean-invariant scalar product of two-wave functionals by 
\beq
(B,C) := \sum_{j=1}^{N_b} \sum_{k=1}^{N_c} b^*_j c_k Z[ g_k- f_j ] .
\label{b.2}
\eeq
This becomes a Hilbert space inner product by identifying 
vectors whose difference has zero norm and adding convergent 
sequences of finite sums.  We call this space the Euclidean Hilbert space.

Reflection positivity can be used to define a second Hilbert space.
Vectors are represented by wave functionals of the form (\ref{b.1})
where the test functions $f_j(\x)$, $g_k(\x)$ are restricted to have
support for positive Euclidean times.  We call these test functions
positive-time test functions.  We define the physical scalar product
of two such wave functionals by
\beq
\langle B \vert C \rangle := 
\sum_{j=1}^{N_b} \sum_{k=1}^{N_c} 
b^*_j c_k Z[ g_k- \Theta f_j ] .
\label{b.3}
\eeq
As in the Euclidean case, this becomes a Hilbert space inner product
by identifying vectors whose difference has zero norm and adding
convergent sequences of finite sums.  We will refer to the resulting
Hilbert space as the physical Hilbert space.  Reflection positivity 
is equivalent to the requirement that 
\beq
\langle B \vert B \rangle \geq 0 .
\eeq

\section{Poincar\'e Lie Algebra} 

Note that the determinant of the $2\times 2$ matrices 
\beq
X=
\left ( 
\begin{array}{cc} 
t-z & x-iy \\
x+iy & t+z 
\end{array} 
\right ) 
\qquad
\X = \left ( 
\begin{array}{cc} 
i \tau-z & x-iy \\
x+iy & i \tau +z 
\end{array} 
\right ) 
\eeq
gives the Lorentz and Euclidean invariant distances.  The determinants
are preserved under the linear transformations $X\to X' = AXB^t$ and
$\X \to \X' = A\X B^t$ where $A$ and $B$ are complex matrices with
determinant 1.  These transformations are generally complex but the
determinants remain real.  It follows that the pair $(A,B)$
equivalently defines both complex Lorentz and complex $O(4)$
transformations.  Real Lorentz transformations have $B=A^*$ while real
$O(4)$ transformations have A and B $\in SU(2)$.  In this section we
use the observation that real $O(4)$ transformations correspond to complex
Lorentz transformations to extract Poincar\'e generators on the
physical Hilbert space.

Finite Euclidean transformations, $T(E,\a)$, act on wave functionals
as follows
\beq
T(E,\a) B[\phi]= \sum_{j=1}^{N_b} b_j e^{i \phi (f_{E,\a,j})},
\eeq
where  $f_{E,\a,j} (\x ) = f_j (E^{-1}(\x -\a ))$ and $E\in O(4)$.  
Since real Euclidean transformations preserve the Euclidean scalar 
product $(\cdot ,\cdot )$,
$T(E,\a)$ is unitary on the Euclidean Hilbert space.

These same transformations, with restrictions on the domains and group
parameters to ensure the positive time support condition is preserved,
are defined on the physical Hilbert space, but the resulting
transformations are not unitary.

For three-dimensional Euclidean transformations, $T(E,\a)$ maps the
physical Hilbert space to the physical Hilbert space in a manner that
preserves the physical Hilbert space scalar product.  This implies that
for space translations and ordinary rotations $T(\E,a)$ is unitary 
on the physical Hilbert space.

Positive Euclidean time translations, $T(I,(\beta, 0))$, $\beta>0$,
map the physical Hilbert space to the physical Hilbert space, however
because of the Euclidean time reversal operator in the physical 
scalar product, Euclidean time
translations are Hermitian, rather than unitary.  It is possible to
use the unitarity of $\Theta$ on the Euclidean Hilbert space along
with reflection positivity\cite{glimm:1981} to show that positive Euclidean
time evolution is a contractive Hermetian semigroup on the physical
Hilbert space.

Rotations in planes that contain the Euclidean time direction do not
generally preserve the positive Euclidean time support constraint.
However, if the test functions are restricted to have support in a cone
with axis of symmetry along the Euclidean time axis that makes an
angle less than $\pi/2$ with the time axis, then rotations in
space-time planes through angles $\rho$ small enough to leave the cone
in the positive-time half plane are defined on this restricted set of
wave functionals.  On this domain and for this restricted set of
angles Euclidean space-time rotations are Hermitian.  They form a local
symmetric semigroup.  What is relevant is that just like one-parameter unitary 
groups and contractive Hermitian semigroups,  local symmetric
semigroups have self-adjoint generators \cite{Klein:1981}
\cite{Klein:1983}\cite{Frohlich:1983kp}.

The result is that on the physical Hilbert space the various 
one-parameter subgroups of the real Euclidean 
transformations have the form
\beq
T(E,\a) \to e^{i  \mathbf{a}\cdot \mathbf{P}},
e^{i {\btheta} \cdot \mathbf{J}  },
e^{-\beta H}, e^{{\brho} \cdot \mathbf{K}}
\eeq
where $H,\mathbf{P},\mathbf{J},\mathbf{K}$ are all self-adjoint operators on
the physical Hilbert space.  It can also be shown by direct calculation
that the infinitesimal generators satisfy the Poincar\'e commutation 
relations.  This is a consequence of the relation between the 
complex Lorentz and complex $O(4)$ groups.

Matrix elements of the generators can be computed by differentiating 
$T(E,\a)$ with respect to the group parameters: 
\beq
\langle B \vert \mathbf{J} \vert C \rangle = 
-i {\partial \over \partial \btheta}
\sum_{j=1}^{N_b} \sum_{k=1}^{N_c} 
b^*_j c_k Z[ g_k- \Theta f_{E(\btheta),0,j} ] _{\vert_{\btheta=0}} 
\eeq
\beq
\langle B \vert \mathbf{P} \vert C \rangle = 
-i {\partial \over \partial \mathbf{a}}
\sum_{j=1}^{N_b} \sum_{k=1}^{N_c} 
b^*_j c_k Z[ g_k- \Theta f_{I,\mathbf{a},j} ] _{\vert_{\mathbf{a}=0}} 
\eeq
\beq
\langle B \vert {H} \vert C \rangle = 
- {\partial \over \partial \beta}
\sum_{j=1}^{N_b} \sum_{k=1}^{N_c} 
b^*_j c_k Z[ g_k- \Theta f_{I,(\beta,0),j} ] _{\vert_{\beta=0}} 
\eeq
\beq
\langle B \vert \mathbf{K} \vert C \rangle = 
{\partial \over \partial \brho}
\sum_{j=1}^{N_b} \sum_{k=1}^{N_c} 
b^*_j c_k Z[ g_k- \Theta f_{E(\brho),0,j} ] _{\vert_{\beta=0}} 
\eeq
where $\brho$ is the axis and angle of a rotation in a plane 
containing the Euclidean time direction (it is an imaginary rapidity). 

In this section we have illustrated how the Poincar\'e generators can 
be constructed directly from the Euclidean generating functional 
without using any analytic continuation.

\section{Particles} 

Particles are associated with  eigenstates of the mass Casimir operator 
of the Poincar\'e group with eigenvalues in the point spectrum.
Matrix elements of the square of the mass operator are
\beq
\langle B \vert M^2 \vert C \rangle =
\left ( {\partial^2 \over \partial \beta^2} +
{\partial^2 \over \partial \mathbf{a}^2} \right ) 
\sum_{j=1}^{N_b} \sum_{k=1}^{N_c} 
b^*_j c_k Z[ g_k- \Theta f_{I,(\beta ,\mathbf{a}),j} ]_{ \vert_{\mathbf{a}=\beta=0}} 
\eeq

Since the positive-time wave functionals are dense in the 
physical Hilbert space it is possible to construct an orthonormal basis 
of wave functionals $\{ B_n [\phi]\}$ satisfying
\beq
B_n [\phi] \qquad \langle B_n \vert B_m \rangle = \delta_{mn} .
\eeq

Point eigenstates of the mass operator are normalizable 
solutions of the eigenvalue problem
\beq
\left ((M^2-\lambda^2)  B_\lambda\right ) [\phi] = 0 .
\eeq
In the orthonormal basis $\{ B_n [\phi]\}$ this eigenvalue equation becomes
\beq
B_\lambda [\phi] = \sum_{n} b_n B_n [\phi] 
\qquad
\sum_n \langle B_m \vert M^2 \vert B_n \rangle b_n = \lambda^2 b_m  
\eeq
where the sum is generally infinite. 

States of sharp linear momentum 
and canonical spin can be extracted using translations and rotations.  
Specifically mass-momentum eigenstates are given by
\beq
\langle C \vert B_\lambda (\mathbf{p}) \rangle = 
\int {d^3a \over (2 \pi)^{3/2}}  e^{-i \mathbf{p}\cdot \mathbf{a}}  
\sum_{j=1}^{N_c} \sum_{n}  
b_n c^*_j Z[f_{I,(0,\mathbf{a}),n} - \Theta g_j   ] 
\eeq
which can be normalized so 
\beq
\langle B_\lambda (\mathbf{p}\,' ) \vert B_\lambda (\mathbf{p}) \rangle
= \delta (\mathbf{p}' - \mathbf{p})
\eeq

Similarly, it is possible find simultaneous eigenstates of 
mass, linear momentum and spin using 
\beq
\langle C  \vert 
B_{\lambda,j} (\mathbf{p},\mu ) \rangle := 
\int_{SU(2)} \sum_{\nu=-j}^j 
dR  
\langle C \vert T(R,0) \vert B_\lambda (R^{-1} \mathbf{p}) \rangle 
D^{j*}_{\mu \nu} (R)  
\eeq
where $dR$ is the $SU(2)$ Haar measure.

The wave functional $B_{\lambda,j} (\mathbf{p},\mu )[\phi]$ 
describes a particle of mass $\lambda$, linear momentum $\mathbf{p}$,
spin $j$ and z-component of canonical spin $\mu$.

We remark that if this state is non-degenerate, then it must transform 
irreducibly with respect to the Poincar\'e group.
This means that
\beq
\langle C \vert U(\Lambda ,a) \vert B_{\lambda,j} (\mathbf{p},\mu ) \rangle := 
\sum_{\mu'=-j}^j  \int d\mathbf{p}' 
\langle C \vert B_{\lambda,j}  (\mathbf{p}', \mu')  \rangle
{\cal D}^{\lambda ,j}_{\mathbf{p}'\mu';\mathbf{p},\mu}[\Lambda ,a] 
\label{c.10}
\eeq
where
\beq
{\cal D}^{\lambda ,j}_{\mathbf{p}'\mu';\mathbf{p},\mu}[\Lambda ,a] 
:=
\langle (\lambda ,j) ,\mathbf{p'},\mu' 
\vert U[\Lambda ,a] \vert  (\lambda ,j) ,\mathbf{p},\mu \rangle
\eeq
is the known Wigner fucntion of the Poincar\'e group in the basis  
$\vert  (\lambda ,j) ,\mathbf{p},\mu \rangle$.

As emphasized in the previous sections, all of the calculations were
done using Euclidean Green functions and test functions, with no 
analytic continuation.  Equation (\ref{c.10}) demonstrates how to 
perform finite Poincar\'e transformation on the one-body solutions.

\section{Scattering}

The conventional treatment of scattering problems in quantum field
theory is formulated using the LSZ asymptotic conditions.  These have
the advantage that they can be implemented without solving the
one-body problem, which is non-trivial in field theories.  However,
given one-body solutions it is also possible to formulate scattering
asymptotic conditions using strong limits.  These asymptotic
conditions were given by Hagg and Ruelle
\cite{Haag:1958vt}\cite{Ruelle:1962}, and they are the most natural
generalization of the formulation of scattering that is used in
non-relativistic quantum mechanics.
   
In this work it is useful to use a two Hilbert space formulation 
\cite{Coester:1965} of
Haag-Ruelle scattering theory\cite{simon}\cite{baumgartl:1983}, where
an asymptotic Hilbert space is introduced the formulate the
asymptotic conditions on the scattering states.  All particles appear
as elementary particles in the asymptotic space; the internal
structure (bound-state wave functions) appear in the mapping to the
physical Hilbert space.
  
For a scalar field theory with a mass eigenstate with eigenvalue
$\lambda$, Haag and Ruelle multiply the Fourier transform of the field
by a smooth function $\rho_\lambda(p^2)$ that is $1$ when
$p^2=-\lambda^2$ and vanishes when $-p^2$ is in rest of the mass
spectrum of the system.  The product, $\tilde{\phi}_\rho
(p):=\tilde{\phi}(p) \rho_\lambda (p^2)$, is Fourier transformed back
to configuration space.  The resulting field, $\phi_\rho(x)$, has the
property that it creates a one-body state of mass $\lambda$ out of the
vacuum.  It transforms covariantly, but is no longer local.  While
this is not a free field, it asymptotically looks like a free field,
and it is useful to extract the linear combination of 
$\phi_\rho(x)$ and $\dot{\phi}_\rho(x)$ that asymptotically 
becomes the creation part of the field: 
\beq 
A(f,t):= -i \int \phi_{\rho} (x)
\stackrel{\leftrightarrow}{\partial_0} f (x) d\mathbf{x}
\label{d.1}
\eeq
where $f(x)$ is a positive-energy solution of the Klein Gordon 
equation with mass $\lambda$.
Haag and Ruelle prove that the scattering states of the theory are
given by the limits: 
\beq 
\lim_{t \to \pm \infty} \Vert \vert
\Psi_{\pm} (f_1, \cdots f_n) \rangle - A(f_n,t) \cdots A(f_1,t) \vert
0 \rangle \Vert =0 .
\label{d.2}
\eeq 
To express this in a two Hilbert space
notation we rewrite $A(f,t)$ as
\beq
A(f,t) \vert 0 \rangle = 
e^{i H t} \int  
\underbrace{\left ( 
[H, \hat{\phi}_{\rho} (\mathbf{p})] -
\omega_{\lambda}(\mathbf{p})
\hat{\phi}_\rho(\mathbf{p})  
\right )}_{A(\mathbf{p})} e^{-iHt} \vert 0 \rangle
d\mathbf{p} e^{-i \omega_\lambda (\mathbf{p})t}
\tilde{f}(\mathbf{p})   
d\mathbf{p}.
\label{d.3}
\eeq
It follows that
\beq
A(f_n,t) \cdots A(f_1,t) \vert 0 \rangle = 
e^{iHt} \int  
\underbrace{A(\mathbf{p}_n) \cdots A(\mathbf{p}_1) \vert 0 
\rangle}_{\Phi: \otimes {\cal H}_i \to {\cal H}}
d\mathbf{p}_n \cdots d\mathbf{p}_1 
e^{-i H_0t}  f_n(\mathbf{p}_n)\cdots f_1(\mathbf{p}_1)
\label{d.4}
\eeq
where
\beq
H_0 =\sum_j \omega_{\lambda} (\mathbf{p}_i).
\label{d.5}
\eeq
In this notation equation (\ref{d.2}) has the form 
\beq
\vert \Psi_{\pm} (f_1, \cdots f_n) \rangle= 
\lim_{t \to \infty} e^{iHt} \Phi e^{-i H_0t} \vert \mathbf{f} \rangle =
\Omega_{\pm} \vert \mathbf{f} \rangle .
\label{d.6}
\eeq
This can be expressed in the Euclidean generating functional representation
by replacing $\tilde{\phi}(p) \rho_\lambda (p^2)$ by the 
$B_{\lambda,j}(\mathbf{p},\mu)$ which also creates a one-body state
of mass $\lambda$ out of the vacuum.  The operator $\Phi$ becomes 
\beq
\Phi (\mathbf{p}_n,\mu_n  \cdots \mathbf{p}_1,\mu_1) [\phi]
:=\left ( \prod \left (  
[H, B_{\lambda_k,j_k} (\mathbf{p}_k,\mu_k )] - \omega_{\lambda_k}(\mathbf{p}_k) 
B_{\lambda_k,j_k} (\mathbf{p}_k,\mu_k )  
\right ) \right )  [\phi] 
\label{d.8}
\eeq
where the wave functionals are treated as multiplication operators 
and
\beq
[H, B][\phi]  = 
{\partial \over \partial \beta} 
\sum b_n e^{i \phi (f_{I,(\beta,0),n})}_{\vert_{\beta=0}} .
\eeq
Two Hilbert space wave operators are defined by  
\beq
\vert \Psi_{\pm} (f_1, \cdots f_n) \rangle= 
\lim_{t \to \infty} e^{iHt} \Phi e^{-i H_0t} \vert \mathbf{f} \rangle =
\Omega_{\pm} \vert \mathbf{f} \rangle .
\eeq
The wave operators satisfy
\beq
U[\Lambda ,a] \Omega_{\pm} = \Omega_{\pm} U_f [\Lambda ,a] 
\qquad
\mbox{where}
\qquad
U_f [\Lambda ,a] = \otimes U_{\lambda_k,j_k} [\Lambda ,a] .
\eeq
Since the asymptotic particles transform like free particles with 
physical masses,  this formula can used to compute finite Poincar\'e
transforms of scattering states.
  
\section{Computational considerations}

One of the difficulties with using the generating functional
representation to do 
scattering calculations is that we have no simple means to
construct $e^{iHt}$ on the physical Hilbert space.  This can be 
overcome using a trick.   In non-relativistic scattering 
theory Kato and Birman \cite{simon}\cite{baumgartl:1983} showed that if
\beq
\lim_{t \to \pm \infty} e^{iHt} \Phi e^{-iH_0t} \vert \psi \rangle =
\vert \psi_{\pm} \rangle    
\eeq
then for admissible functions $\chi$ 
\beq
\lim_{t \to \pm \infty} e^{i\chi(H)t} \Phi e^{-i\chi(H_0)t} \vert \psi \rangle =
\vert \psi_{\pm} \rangle .    
\label{e.2}
\eeq
A useful choice of an admissible $\chi$ is $\chi (x) = - e^{-\beta x}$
for $\beta >0$.  If this result is used in (\ref{e.2}) then
we have the alternative representation of the scattering state
\beq
\vert \psi_{\pm} \rangle =   
\lim_{n \to \pm \infty} e^{-in e^{-\beta H}} \Phi 
e^{in e^{-\beta H_0}} \vert \psi \rangle .
\label{e.3}
\eeq
The advantage of this representation is that because $H\geq 0$, the
spectrum of $(e^{-\beta H})$ is in the interval $[0,1]$.  For large
fixed $n$, $e^{-in e^{-\beta H}}$ can be uniformly approximated by a
polynomial in $e^{-\beta H}$.  The advantage is that powers of
$e^{-\beta H}$ can be computed directly using the generating
functional
\beq
\langle C \vert e^{-\beta n H} \vert B \rangle = 
\sum_{j=1}^{N_b} \sum_{k=1}^{N_c} 
b^*_j c_k Z[ g_{I,(n\beta ,0),k}- \Theta f_j ] 
\eeq
without using analytic continuation.

This suggest the following sequence of approximations to compute scattering 
amplitudes.  First use narrow wave packets sharply peaked 
in linear momentum to approximate sharp
momentum transition matrix elements in terms of $S$ matrix elements 
in normalizable states:
\beq
\langle \mathbf{p}_1' ,\mu_1', \cdots , \mathbf{p}_n', \mu_n' \vert  
T \vert \mathbf{p}_1 ,\mu_1,\mathbf{p}_2, \mu_2 \rangle
\approx  
{\langle \Psi_{f}' \vert S \vert \Psi_{f} \rangle
-
\delta_{ab}\langle \Psi_{f}' \vert \Psi_{f} \rangle
\over 2 \pi i 
\langle \Psi_{f}' \vert \delta (E_+-E_-) \vert \Psi_{f} \rangle } .
\eeq
Next approximate $\langle \Psi_{f}' \vert S \vert \Psi_{f} \rangle =
\langle \Psi_{f+}' \vert \Psi_{f-} \rangle$ using (\ref{e.3}) for large
enough $n$.   This step also involves solving the one-body 
problem for each asymptotic particle in the initial and final states:
\beq
\langle \Psi_{f+} \vert \Psi_{f-} \rangle
\approx \langle \Psi_{f} \vert e^{-in e^{-\beta H_f}} 
\Phi^{\dagger}  e^{2in {e^{-\beta H}}} \Phi e^{-in e^{-\beta H_f}} 
\vert \Psi_{f} \rangle .
\label{e.6}
\eeq
Next, {\it after fixing} $n$, approximate $e^{i2nx}$ on $x\in [0,1]$ by a
polynomial in $x$, which gives  
\beq
e^{2in e^{-\beta H}} \approx \sum c_m(n) (e^{-\beta m H}) .
\eeq
Taken together these approximations, when performed in the correct order, 
provide a means to compute on-shell transition matrix elements using purely 
Euclidean methods.

\section{Test of approximations} 

A mathematically controlled approximation is not automatically useful
in all applications. To test the suggested sequence of approximations
at the relevant GeV scale we consider a simple model based on a
separable potential
\[
H = {\mathbf{k}^2 \over m} - \vert g \rangle \lambda \langle g \vert 
\qquad
\langle \mathbf{k} \vert g \rangle = {1 \over m_{\pi}^2 + \mathbf{k}^2}.
\]
This is an exactly solvable model; a first test of the proposed
method is to calculate scattering amplitudes in this model using
matrix elements of $e^{-\beta H}$ in normalizable states.

\begin{figure}
\begin{minipage}[t]{6.3cm}
\begin{center}
\includegraphics[width=6.0cm,clip]{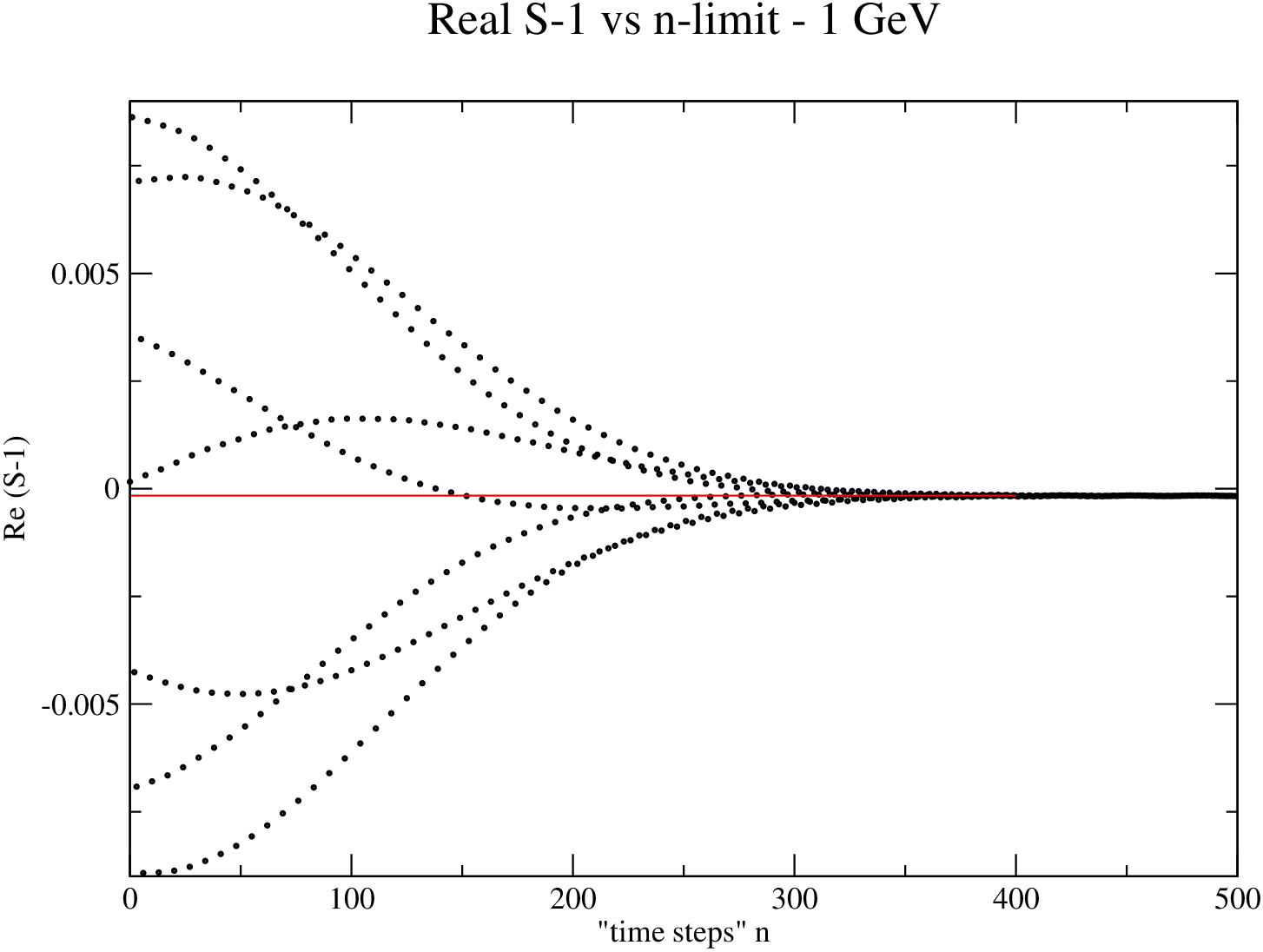}
\caption[Short caption for figure 1]{\label{labelFig1} 
Real part of S compared to Kato Birman approximation as function of
$n$.
}
\end{center}
\label{fig.1}
\end{minipage}
\hfill
\begin{minipage}[t]{6.3cm}
\begin{center}
\includegraphics[width=6.0cm,clip]{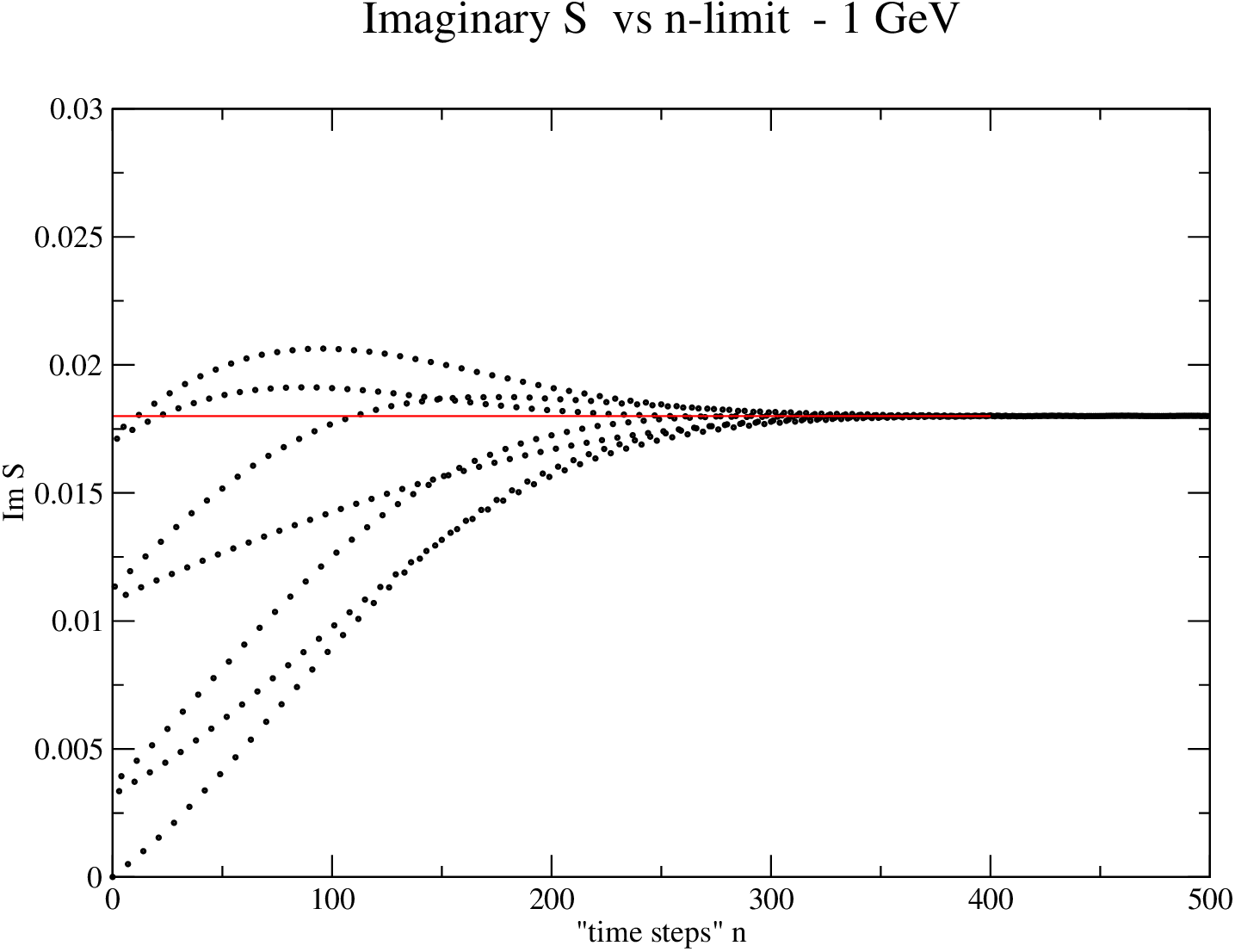}
\caption[Short caption for figure 2]{\label{labelFig2} 
Imaginary part of S compared to Kato Birman approximation as function of
$n$.}
\end{center}
\end{minipage}
\label{fig.2}
\end{figure}
\begin{center}
{\bf
Table 1: Degree 300 polynomial compared to $e^{-inx}$, $n=220$ 
}
\end{center} 
\begin{center}
\begin{tabular}{lll}
\hline
 $x$ & $\Delta \cos (nx)$ & $\Delta \sin (nx)$ \\
\hline
	0    & $ 4.44089\times10^{-16}$	&     $	8.32667\times10^{-15} $ \\	
	0.1  & $ 2.35367\times10^{-14}$	&     $	1.46966\times10^{-14} $ \\	
	0.2  & $ 5.55112\times10^{-16}$	&     $	3.6797\phantom{0}\times10^{-14} $ \\	
	0.3  & $ 3.84137\times10^{-14}$	&     $	1.80689\times10^{-14} $ \\	
	0.4  & $ 1.72085\times10^{-14}$	&     $	1.32672\times10^{-14} $ \\	
	0.5  & $ 2.77556\times10^{-15}$	&     $	2.93793\times10^{-14} $ \\	
	0.6  & $ 6.66134\times10^{-16}$	&     $	3.33344\times10^{-14} $ \\	
	0.7  & $ 8.54872\times10^{-15}$	&     $	2.50355\times10^{-14} $	\\	
	0.8  & $ 1.02141\times10^{-14}$	&     $	1.35447\times10^{-14} $	\\	
	0.9  & $ 1.22125\times10^{-15}$ &     $	2.72282\times10^{-14} $	\\	
	1    & $ 4.88498\times10^{-15}$	&     $	6.61415\times10^{-14} $	\\	
\hline
\end{tabular}	
\end{center}       	       			       

\begin{figure}
\begin{minipage}[t]{6.3cm}
\begin{center}
\includegraphics[width=6.0cm,clip]{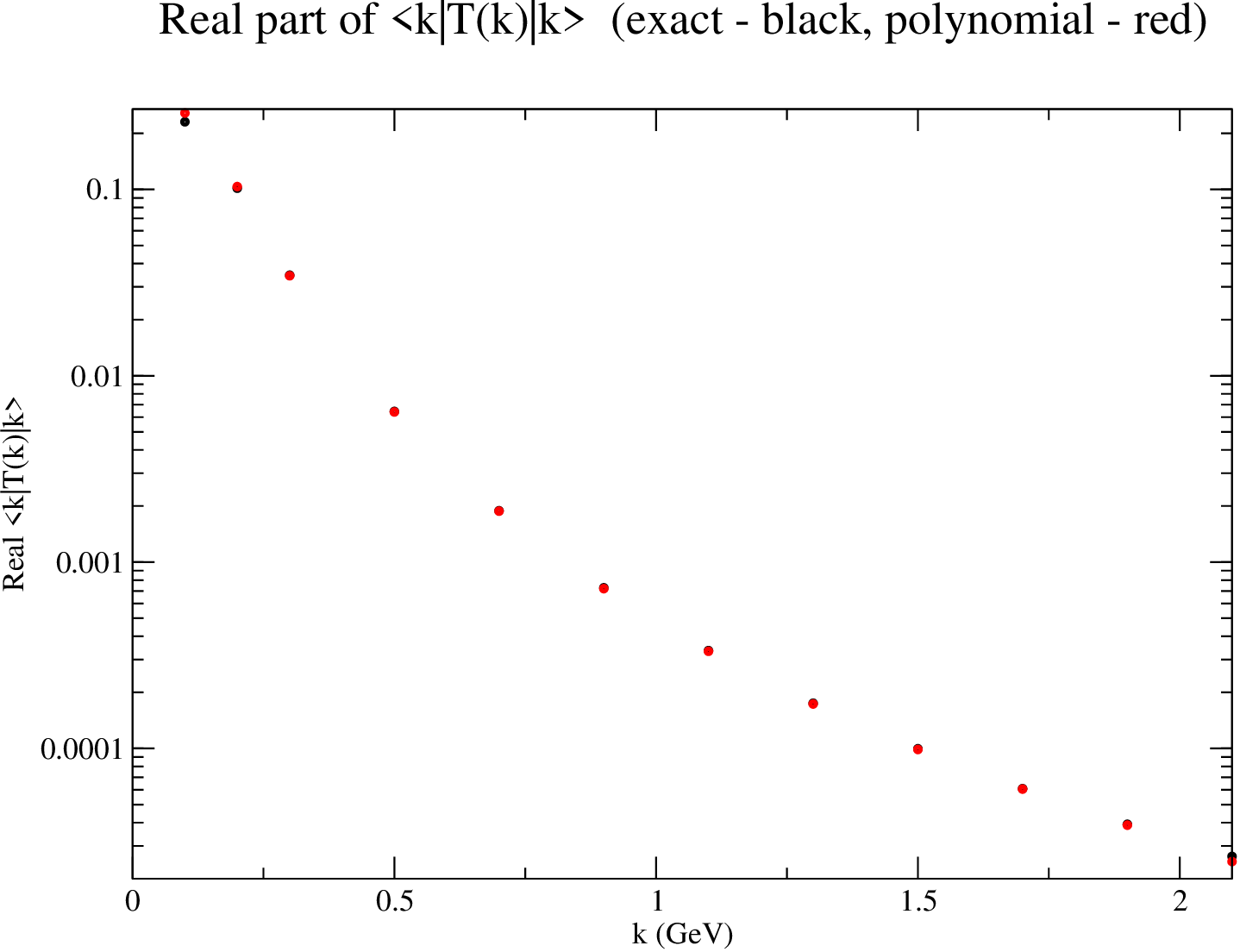}
\caption[Short caption for figure 3]{\label{labelFig3} 
Calculated compared to approximate calculations for the real part of
transition matrix element for different energies.
}
\end{center}
\label{fig.3}
\end{minipage}
\hfill
\begin{minipage}[t]{6.3cm}
\begin{center}
\includegraphics[width=6.0cm,clip]{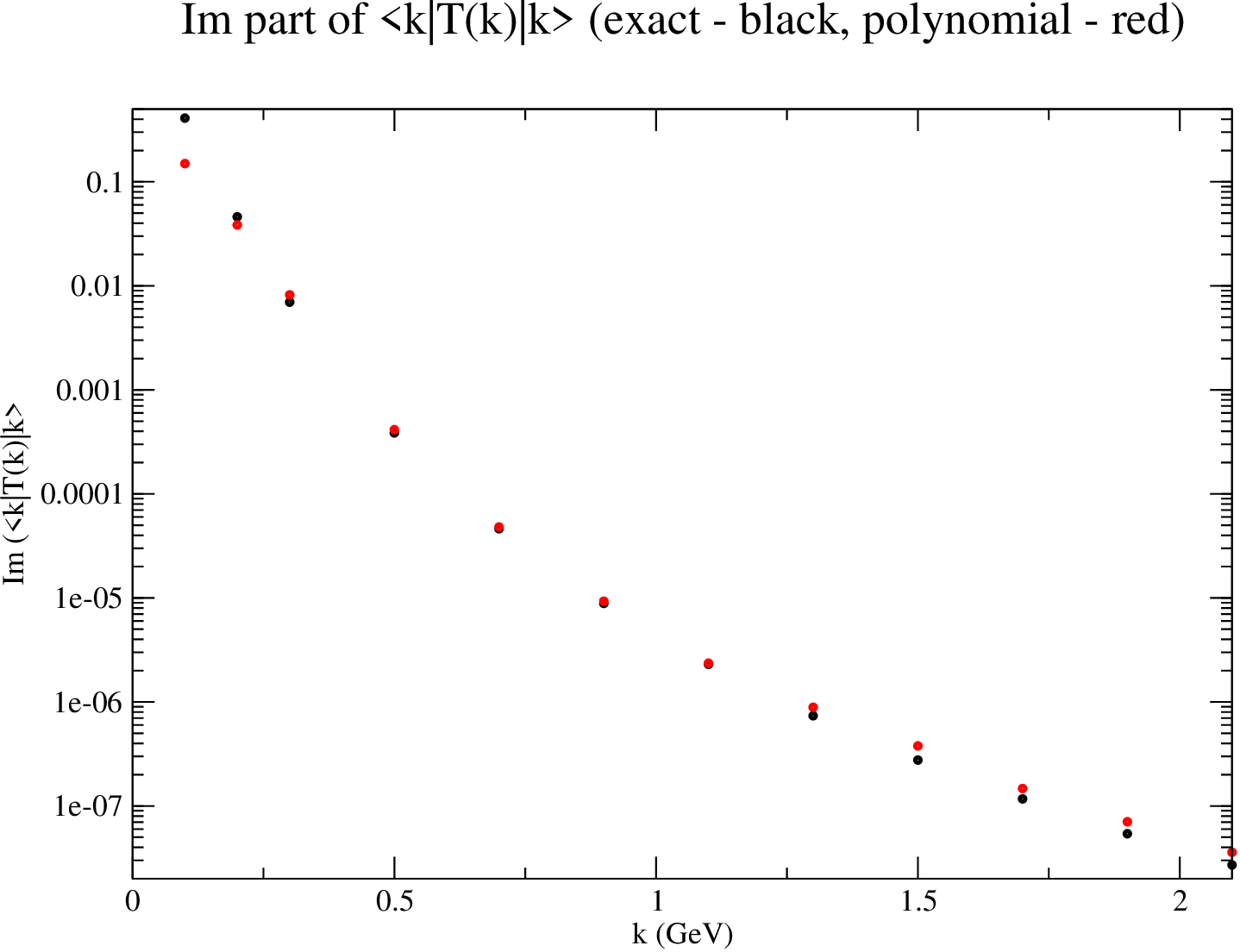}
\caption[Short caption for figure 4]{\label{labelFig4}
Calculated compared to approximate calculations for the real part of
transition matrix element for different energies.}
\end{center}
\end{minipage}
\label{fig.4}
\end{figure}

In this model sharp $T$ matrix elements can be calculated with an error 
of approximately 1\%  using wave packets whose momentum widths are about 
1/10 of the cm momentum.  Figures 1 and 2 show the convergence 
of the real and imaginary parts of the $S$ matrix evaluated in these
wave packets as a function of $n$ in equation (\ref{e.6}).
Values of $n$ between 200-300 are adequate in this model.
$\beta$ is a parameter that can be adjusted to improve the 
convergence.  
Polynomial approximations to $e^{2ine^{-\beta H}}$ are performed 
using Chebyshev expansions:  
\beq 
e^{inx}  \approx {1 \over 2} c_0 T_0 (x) + \sum_{k=1}^N c_k T_k (x) 
\qquad
c_j = {2 \over N+1} \sum_{k=1}^N e^{in ( \cos({2k-1 \over N+1}{\pi
\over 2}))} \cos(j {2k-1 \over N+1}{\pi \over 2}). 
\eeq
Polynomials of degree approximately 300 agree with $e^{2ine^{-\beta
H}}$ uniformly to better than 13 significant figures.

Typical results are shown in table 1.  Figures 3 and 4 compare the
exact value of the real and imaginary parts of the sharp momentum
transition matrix to the calculated values for momenta up to 2 GeV.
For most values of $k$ the exact and approximate values cannot
be distinguished.

These results suggest that it may be feasible to use this method to
formulate relativistic few-body models.  The open problems that have
not been addressed in this preliminary work involve finding model
Green functions or generating functionals satisfying the required
conditions.  Reflection positivity appears to be a fairly restrictive
condition that requires additional study.  The toy model discussed 
above did not require solutions of the one-body problem.  How
approximate solutions of the one-body problem are affected by the other 
approximations also requires further study.

This work supported in part by the U.S. Department of Energy, under 
contract DE-FG02-86ER40286.


\end{document}